# Description of electromagnetic fields in inhomogeneous accelerating sections. II Fields in the regular part

## M.I. Ayzatsky

National Science Center "Kharkiv Institute of Physics and Technology" (NSC KIPT), 61108, Kharkiv, Ukraine

E-mail: mykola.aizatsky@gmail.com

In this work we present the results of calculation of the electric field distribution in the inhomogeneous accelerating section on the base of generalized coupled modes theory. It was shown that the proposed coupled differential equations correctly describe electromagnetic field distribution in the regular part of the section under consideration, which is an analogue of the one being developed at CERN.

## 1. INTRODUCTION

Recently it was proposed to use a uniform basis for description of non-periodic structured waveguides and a new generalization of the theory of coupled modes was constructed [1,2]. One of the positive features of proposed coupled modes theory is the simple and clear procedure for taking into account the beam loading and the easy transition to the case of a homogeneous waveguide [3,4]. It was shown that an infinitive system of equations can be reduced and the single mode representation describes the electromagnetic fields with small errors in the regular part of the accelerating section. In the couplers the accuracy is worse, but it is small enough to be used in calculations [2,5].

Within the single-mode approximation, the fields are represented as the sum of two components, one of which is associated with the forward eigenwave and the other with the backward eigenwave. It was shown that the second component is not a backward wave [2,5]. This is not an obvious result. There are studies taking into account the connection of the forward and backward wave components in the generalized WKB consideration of wave propagation in inhomogeneous media. Attention was paid to the calculation of the reflection coefficient and were not interested in the field distribution (see, for example, [6]and [7]). In our case, the coupling of waves due to the presence of distributed coupling leads to the fact that the characteristics of the backward component are determined by the characteristics of the forward component, which is significantly greater than the first. This results in the phase distribution of the backward wave component having an increasing character, similar to the phase distribution of the forward wave component. As for the field pattern in the case of coupling of waves in inhomogeneous media, it requires additional consideration.

The calculation results showed that the coupling coefficients in the equations of coupled mode theory have a complex dependence on the coordinate $z$. Therefore, accurate numerical methods are required to solve the coupling equations. This paper presents the results of a study of the accuracy of solving the coupling equations along the regular part of the acceleration section, which is a model of the section being developed at CERN [8]. Calculating the solutions of the coupling equations in the coupler region is more complex and requires the development of a special procedure.

## 2. MAIN EQUATIONS

Electromagnetic fields in a non-periodic structured waveguide with the ideal metal walls can be represented in the form of such series ([1])

$$\vec{H}(\vec{r}) = \sum_{s=-\infty}^{s=\infty} C_s(z)\vec{H}_s^{(e,z)}(\vec{r}),$$

$$\vec{E}(\vec{r}) = \sum_{s=-\infty}^{s=\infty} C_s(z)\vec{E}_s^{(e,z)}(\vec{r}) + \frac{\vec{j}_z}{i\omega\varepsilon_0\varepsilon}, \tag{1}$$

where $\vec{E}_s^{(e,z)}(\vec{r}), \vec{H}_s^{(e,z)}(\vec{r})$ are modified eigen vector functions obtained by generalizing the eigen $\vec{E}_s^{(e)}, \vec{H}_s^{(e)}$ vectors of a homogeneous waveguide by special continuation of the geometric parameters ([1,2]). The eigen waves of a homogeneous waveguide we present as $\left(\vec{E}_s, \vec{H}_s\right) = \left(\vec{E}_s^{(e)} \vec{H}_s^{(e)}\right)\exp\left(\gamma_s z\right)$, where ($\vec{E}_s^{(e)} \vec{H}_s^{(e)}$) are the periodic functions of the z-coordinate. Under such choice of the basis functions, the coefficients $C_s(z)$ include an exponential dependence on the z-coordinate.

Electromagnetic fields (1) must obey the Maxwell's equations. To satisfy this condition $C_s(z)$ must be solutions of such a coupled system of differential equations [1,2]

$$\frac{dC_s}{dz} - \gamma_s^{(e,z)}C_s + \frac{1}{2N_s^{(e,z)}}\frac{dN_s^{(e,z)}}{dz}C_s + \sum_{s'=-\infty}^{\infty}C_{s'}U_{s',s}^{(z)} = \frac{1}{N_s^{(e,z)}}\int\limits_{S_\perp^{(z)}}\vec{j}\vec{E}_{-s}^{(e,z)}dS, \tag{2}$$

where

$$U_{k',k}^{(z)} = \frac{1}{2N_k^{(z)}} \sum_i \frac{dg_i^{(z)}}{dz} \int\limits_{S_1^{(z)}(z)} \left\{ \left[ \frac{\partial \vec{E}_{k'}^{(e,z)}}{\partial g_i^{(z)}} \vec{H}_k^{(e,z)} \right] + \left[ \frac{\partial \vec{E}_k^{(e,z)}}{\partial g_i^{(z)}} \vec{H}_{k'}^{(e,z)} \right] - \left[ \vec{E}_k^{(e,z)} \frac{\partial \vec{H}_{k'}^{(e,z)}}{\partial g_i^{(z)}} \right] - \left[ \vec{E}_{k'}^{(e,z)} \frac{\partial \vec{H}_k^{(e,z)}}{\partial g_i^{(z)}} \right] \right\} \vec{e}_z dS, \quad (3)$$

$g_i^{(z)}(z)$ -generalized geometrical parameters, $\gamma_s^{(e,z)}(z)$ - generalized wavenumber and

$$N_s^{(e,z)} = \int\limits_{S_1^{(z)}(z)} \left\{ \left[ \vec{E}_s^{(e,z)} \vec{H}_{-s}^{(e,z)} \right] - \left[ \vec{E}_{-s}^{(e,z)} \vec{H}_s^{(e,z)} \right] \right\} \vec{e}_z dS = \begin{cases} N_s^{(z)}, & s > 0, \\ -N_s^{(z)}, & s < 0. \end{cases} \quad (4)$$

The accelerating section under consideration, which is an analogue of the one being developed at CERN [8], consists of 26 regular sells and two "electric" couplers [5].

We will consider axisymmetric TH (E) electromagnetic fields in the model section without current ( $\vec{j} = 0$ ). In this case $\vec{E}_k^{(e,z)} = \vec{e}_r E_{r,k}^{(e,z)} + \vec{e}_z E_{z,k}^{(e,z)}$ and $\vec{H}_k^{(e,z)} = \vec{e}_\varphi H_{\varphi,k}^{(e,z)}$ .

In the single mode approach, when the representation transforms into

$$\vec{E}_1(\vec{r}) = \vec{E}_1^+(\vec{r}) + \vec{E}_1^-(\vec{r}) = C_1(z) \vec{E}_1^{(e,z)}(\vec{r}) + C_{-1}(z) \vec{E}_{-1}^{(e,z)}(\vec{r}), \quad (5)$$

the coupled system (2) is written as

$$\frac{dC_1}{dz} - \gamma_1^{(e,z)} C_1 + \frac{1}{2N_s^{(z)}} \frac{dN_s^{(z)}}{dz} C_1 + C_1 U_{1,1} + C_{-1} U_{-1,1} = 0$$
$$\frac{dC_{-1}}{dz} + \gamma_1^{(e,z)} C_{-1} + \frac{1}{2N_s^{(z)}} \frac{dN_s^{(z)}}{dz} C_2 + C_{-1} U_{-1,-1} + C_1 U_{1,-1} = 0 \quad . \quad (6)$$

Taking into account that $U_{-1,-1} = -U_{1,1}$ and introducing new functions $\tilde{C}_{\pm 1}$,

$$C_{\pm 1} = \sqrt{\frac{N_1^{(z)}(z_0)}{N_1^{(z)}(z)}} \exp\left( \pm \int\limits_{z_0}^z \left( \gamma_1^{(e,z)} - U_{1,1} \right) dz' \right) \tilde{C}_{\pm 1}, \quad (7)$$

the system (6) takes the form

$$\frac{d\tilde{C}_1}{dz} = -\exp\left( -2\int\limits_{z_0}^z \left( \gamma_1^{(e,z)} - U_{1,1} \right) dz' \right) U_{-1,1} \tilde{C}_{-1},$$
$$\frac{d\tilde{C}_{-1}}{dz} = -\exp\left( 2\int\limits_{z_0}^z \left( \gamma_1^{(e,z)} - U_{1,1} \right) dz' \right) U_{1,-1} \tilde{C}_1. \quad (8)$$

To apply the system (8) to study the field distribution in the accelerating section we need to know not only the values of functions $\gamma_1^{(e,z)}, U_{1,1}, U_{-1,1}, U_{1,-1}$, but also the values of functions $C_{\pm 1}(z)$ at some point $z = z_0$ .

The often-used model for study inhomogeneous electrodynamic objects is the infinitive model, in which the object is placed between two semi-infinitive media that are homogeneous at $z \to \pm\infty$ . In this case the boundary condition at the infinity (no wave coming from $+\infty$ ) can be used and the math problem become definite. The regular part of the accelerating section (without the couplers) can be studied in such way if we replace the couplers with semi-infinite periodic waveguides.

Since we wanted to evaluate the accuracy of the description of the field distribution along regular part of the section by the coupled system (8), we used a different approach. We calculated the field distribution $\vec{E}^E, \vec{H}^E$ on the base of the CASCIE code (Code for Accelerating Structures - Coupled Integral Equations) [9,10], expanded them into the serios (1) by direct integrating of the fields [2,5]

$$C_s^E(z) = \frac{1}{N_s^{(z)}} \int\limits_{S_1^{(z)}(z)} \left( \left[ \vec{E}^E \vec{H}_{-s}^{(e,z)} \right] - \left[ \vec{E}_{-s}^{(e,z)} \vec{H}^E \right] \right) \vec{e}_z dS, \quad (9)$$

found the values of complex amplitudes $C_s^E(z)$ at $z = 3D/2$ (the middle of the second period[1]) and used them as boundary values for solving system (8) $C_{\pm 1}(z_0 = 3D/2) = C_\pm^E(z_0)$ . By solving system (8), we obtain the functions $C_{\pm 1}(z)$ and then, using (5), we calculate the field distribution. By comparing this distribution with that obtained using the CASCIE code, we can evaluate the accuracy of the proposed approach.

## 3. CALCULATION RESULTS

Modification of the eigenmodes of a homogeneous waveguide for use in expansion (1) is accomplished by a special choice of the geometric parameters of the base waveguide at each value of longitudinal coordinate $z$ [1,2]. This transforms the piecewise task into the continues one that can be solved using differential equations.

---

[1] For definitions of symbols and geometric parameters, see [5]

In the accelerating section under consideration, four geometric parameters change along the regular part: the thickness of the disks and the length of the resonators, the radii of the holes and resonators [5,8].

Under calculation of generalized eigen vectors $\vec{E}_s^{(e,z)}, \vec{H}_s^{(e,z)}$ it is necessary to use generalized geometrical parameters $g_i^{(z)}(z)$, which are selected taking into account the following rules.

To satisfy the boundary conditions, it is necessary to keep the generalized disk thickness and the generalized hole radius constant along each disk, while we can vary the generalized resonator length and the generalized resonator radius. Along each resonator, it is necessary to keep the generalized resonator length and its generalized radius constant, while we can vary the generalized disk thickness and the generalized hole radius (see Figure 1).

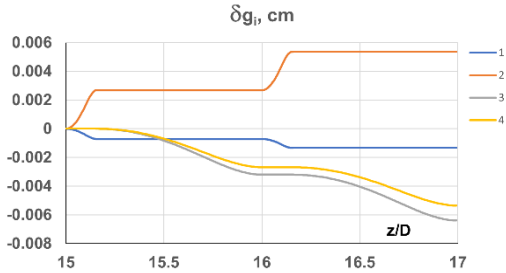

*Figure 1 Increment of generalized geometrical parameters along two cells: 1 – generalized cell radius, 2- generalized cell length, 3 - generalized hole radius, 4 – generalized disk thickness*

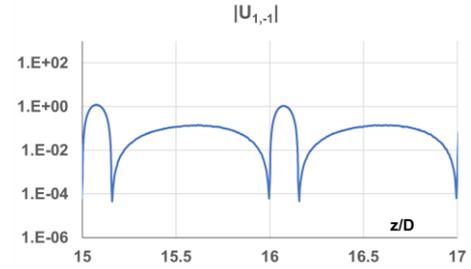

*Figure 2 Modulus of the coupling coefficient $U_{1,-1}^{(z)}$*

Since the derivatives of the generalized geometric parameters $g_i^{(z)}(z)$ differ greatly in the regions of disks and resonators, and the coupling coefficients are proportional to $dg_i^{(z)}/dz$, $U_{k',k}^{(z)}$ will also have greatly different values in these regions (see Figure 2). At the edges of the disk and resonator regions $dg_i^{(z)}/dz \to 0$, therefore, $U_{k',k}^{(z)}$ also tends to zero. Note that in each region the sum for the coupling coefficients $U_{k',k}^{(z)}$ contains only two terms.

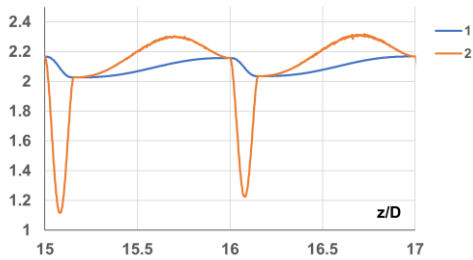

*Figure 3 Dependences on $z$ of the local logitudinal wavenumber: 1- $\gamma_1^{(z)}$, 2 - $\gamma_1^{(z)} - U_{1,1}$*

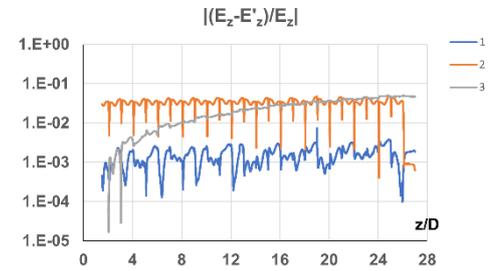

*Figure 4 Modulus of relative error in the representation of the longitudinal electric field at $r = 0$: $E_z^E$ - calculated using the CASCIE code, 1 - $E_z' = E_{1,z}$ (see (5)), 2 - $E_z' = E_{1,z}^+$, 3- $E_z' = E_{1,z}^{(1)}$, $E_{1,z}^{(1)}$ - the first approximation*

As can be seen from (7), that due to the coupling, the generalized longitudinal wave vector $\gamma_1^{(e,z)}$ changes by the value $(-U_{1,1})$. This addition is not small, but has an alternating character (Figure 3) which reduces its influence on the distribution of phases across the entire section.

To solve the system (8), each period was divided into $M$ equal intervals $h = D/M$. The integral in (8) was calculated using Simpson's method and the system of differential equations (8) was solved using Runge-Kutta method. Since in Simpson's method 3 points are required to perform one calculation step, the integral can be calculated with a step $2h$. This led to fact, that we can calculate $\tilde{C}_1$ and $\tilde{C}_{-1}$ with a step $4h$.

Using such approach for solving the system (8), we obtained the functions $C_{\pm 1}(z)$ and then, using (5), we calculated the field distribution. Comparing this distribution with that obtained using the CASCIE code is presented on Figure 4 ($M = 1000$). It can be seen that accuracy of field calculation on the base of system (8) is from 0.1% to 1%. Changing the number of divisions $M$ from 500 to 2000 does not increase the accuracy. This is due to the fact that the accuracy is determined not only by the errors of the difference scheme, but also by the errors in calculating the coupling coefficients $U_{k',k}^{(z)}$ (see (3)) which are difficult to control. Using only the first part in the representation of field (5) $\vec{E}_1(\vec{r}) = \vec{E}_1^+(\vec{r}) = C_1(z)\vec{E}_1^{(e,z)}(\vec{r})$ gives an error of about 5% (see Figure 4).

The most unexpected result of study the representation of the field distribution in the non-periodic structured waveguides was the fact that the part, connected with the backward field, is not the backward wave [2]. Calculation on the base of the system (8) confirms this result. On the Figure 5 and Figure 6 results of calculations of the spatial distributions of the amplitudes $C_1$ and $C_{-1}$ based on different approaches are presented. It can be seen that results are in good agreement and that $C_{-1}$ is not the backward wave as its phase increase along the axis $z$.

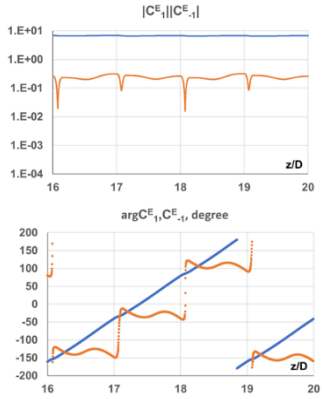 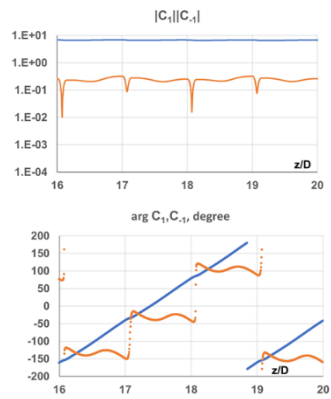 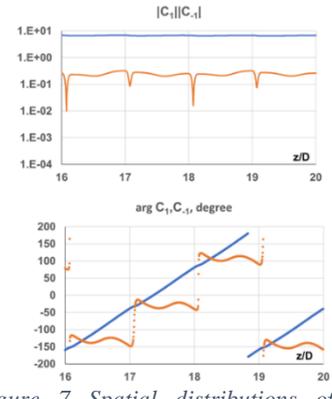

*Figure 5 Spatial distributions of the modulus and arguments of amplitudes $C_1^E$ (blue) and $C_{-1}^E$ (red) calculated based on formular (9).*

*Figure 6 Spatial distributions of the modulus and arguments of amplitudes $C_1$ (blue) and $C_{-1}$ (red) calculated based on the system (8).*

*Figure 7 Spatial distributions of the modulus and arguments of amplitudes $C_1$ (blue) and $C_{-1}$ (red) calculated based on the system (8) in the first approximation ( $U_{-1,1}=0$ ).*

It was supposed that in the model of coupling waves part of field that is connected with the backward field, is the backward wave. "In other words, as a result of the inhomogeneity of the medium, a wave of the zeroth approximation undergoes reflection at every point of space and gives rise to a wave of the first approximation propagating in the opposite direction" [6]. In our case this is not correct. To calculate $C_{-1}$ in the first approximation it is necessary to set $\tilde{C}_1 = const$ ($U_{-1,1}=0$) . [6]. Results of such calculations are shown in Figure 7. It can be seen that the phase distribution didn't change its character. A comparison of the results presented in the Figure 5 and Figure 7 shows that the first approximation yields distributions that differ from those obtained using the CASCIE code by more than 4% (see Figure 7, line 3). The same error will be in the case of using only the first part in the representation of field (5).

## 4. CONCLUSIONS

New generalization of the theory of coupled modes, proposed in [1], gives possibility to describe nonuniform accelerating structures. Based on a set of eigen waves of a homogeneous periodic waveguide, a new basis of vector functions is introduced that takes into account the non-periodicity of the waveguide. Representing the total field as the sum of these functions with unknown scalar coefficients, an infinitive system of coupled equations that determines the dependence of these coefficients on the longitudinal coordinate can be obtained. In some cases it can be reduce and only the single mode approximation can be considered.

In this work we showed that using the system of coupled equation for description of the field distribution along the regular part of the accelerating section gives the error of calculation less than 1%.

Within the framework of single wave approximation, the fields are represented as sum of two components, one of which is associated with the forward eigen wave, and the second with the backward eigen wave. But this second component is not a backward wave. This result is confirmed by calculations on the base of coupling equations.